\documentclass[
]{ceurart}

\usepackage{listings}
\begin{document}

\copyrightyear{2024}
\copyrightclause{Copyright for this paper by its authors.
  Use permitted under Creative Commons License Attribution 4.0
  International (CC BY 4.0).}

\conference{Woodstock'24: Symposium on the irreproducible science,
  June 07--11, 2024, Woodstock, NY}

\title{Integrating SPARQL and LLMs for Question Answering over Scholarly Data Sources}


\author[1]{Fomubad Borista Fondi}[%
orcid=0009-0005-9448-7722,
email=borista.fomubad@facsciences-uy1.cm,
]

\author[1,2]{Azanzi Jiomekong}[%
orcid=0000-0002-8005-2067,
email=fidel.jiomekong@facsciences-uy1.cm
]

\author[3]{Gaoussou Camara}[%
orcid=0000-0001-5223-2282,
email=gaoussou.camara@uadb.edu.sn
]


\address[1]{Department of Computer Science, University of Yaounde I, Yaounde, Cameroon}
\address[2]{TIB – Leibniz Information Centre for Science and Technology, Hannover, Germany}
\address[3]{Unité de Formation et de Recherche en Sciences Appliquées et des TIC, Université Alioune Diop de Bambey du Sénégal }


\maketitle

\begin{abstract}
The Scholarly Hybrid Question Answering over Linked Data (QALD) Challenge at the International Semantic Web Conference (ISWC) 2024 focuses on Question Answering (QA) over diverse scholarly sources: DBLP, SemOpenAlex, and Wikipedia-based texts. This paper describes a methodology that combines SPARQL queries, divide and conquer algorithms, and a pre-trained extractive question answering model. It starts with SPARQL queries to gather data, then applies divide and conquer to manage various question types and sources, and uses the model to handle personal author questions. The approach, evaluated with Exact Match and F-score metrics, shows promise for improving QA accuracy and efficiency in scholarly contexts.

\end{abstract}

\section{Introduction}
\label{introduction}

The Scholarly Hybrid Question Answering over Linked Data (QALD) aims to answer hybrid questions in scholarly publications provided in natural language~\cite{qald2023}. The challenge focuses on Question Answering over Linked Data (QALD) and has been hosted at the International Semantic Web Conference (ISWC) 2024~\cite{iswc2024} since 2023. The 2024 edition is devoted to the development of question answering (QA) systems capable of integrating and querying information from three distinct but interconnected sources: DBLP Knowledge Graph\footnote{\url{https://dblp-april24.skynet.coypu.org/sparql}}, SemOpenAlex Knowledge Graph\footnote{\url{https://semoa.skynet.coypu.org/sparql}}, and Wikipedia-based texts\footnote{\url{https://drive.google.com/file/d/1ISxvb4q1TxcYRDWlyG-KalInSOeZqpyI/view?usp=drive_link}}.

\begin{enumerate}
    \item \textbf{DBLP Knowledge Graph} is a comprehensive dataset documenting research publications, authors, and affiliations.
    \item \textbf{SemOpenAlex Knowledge Graph} is an extensive KG containing detailed information about authors, institutions, and publications.
    \item \textbf{Wikipedia-Based Scholarly Text} is composed of textual data derived from Wikipedia, offering supplementary information on scholarly topics.
\end{enumerate}

The primary objective of this paper is to describe the methodology employed in addressing the Scholarly Hybrid QALD Challenge. This includes detailing the integration of SPARQL queries across different KGs~\cite{AjVtBf2019AtP}, the application of divide-and-conquer algorithms~\cite{divide_conquer_ref}, and the utilization of BERT~\cite{devlin2018bert} to improve response accuracy. To assess this methodology, the dataset provided by the organisers was used. This dataset was composed of training set and test set. The training set was composed of 5000 questions along with their answers, while the test set was composed of 702 questions. The approach proposed in this paper shows promising results for this challenge.

The rest of the paper is organised as follows: Section \ref{methodology} is the detailed methodology used, the Section \ref{results} presents the results and the Section \ref{conclusion} conclude this work.


\section{Methodology}
\label{methodology}
\begin{figure}[h!]
    \centering
    \includegraphics[width=1.0\linewidth]{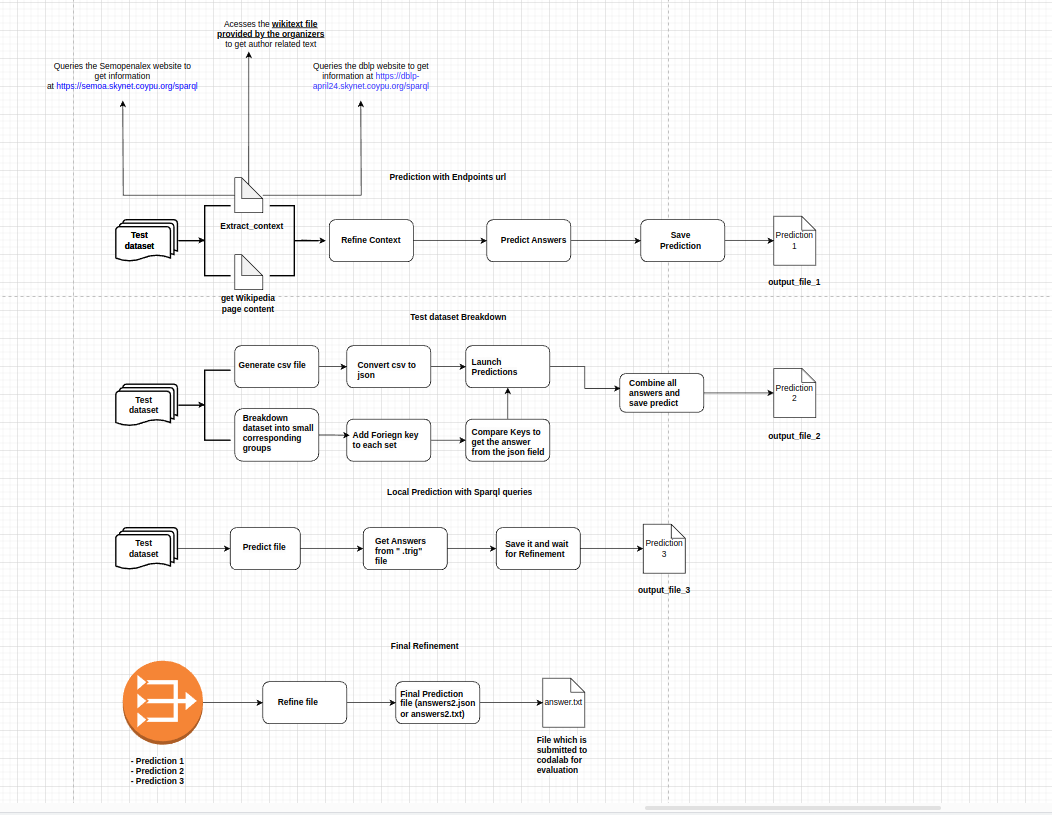}
    \caption{Methodology pipeline for the Scholarly Hybrid QALD Challenge.}
    \label{fig:pipeline-2}
\end{figure}

To address the Scholarly Hybrid Question Answering over Linked Data (QALD) Challenge, we adopted a multi-step approach combining natural language processing techniques for data processing, SPARQL queries, divide and conquer algorithms, and LLM-based predictions. This methodology is designed to efficiently handle the complexity of integrating information from multiple sources and producing accurate answers for a given set of questions. Fig. \ref{fig:pipeline-2} provides an overview of the methodology pipeline used in this work. This figure illustrates the main steps involved in processing the data, executing queries, applying LLM-based predictions, generating answers, and refining them.

\subsection{Data Processing and Query Execution}
\label{dataProcessing}
The process began with executing a general script containing SPARQL queries against SemOpenAlex for both authors and institutions. This step involved querying the \texttt{semoa\_authors.trig} file and the \texttt{institution-semopenalex.trig} dataset from October 2023 locally. The query execution took approximately 62-65 hours to complete due to the size and complexity of the datasets.

\begin{figure}[h!]
    \centering
    \includegraphics[width=1.0\linewidth]{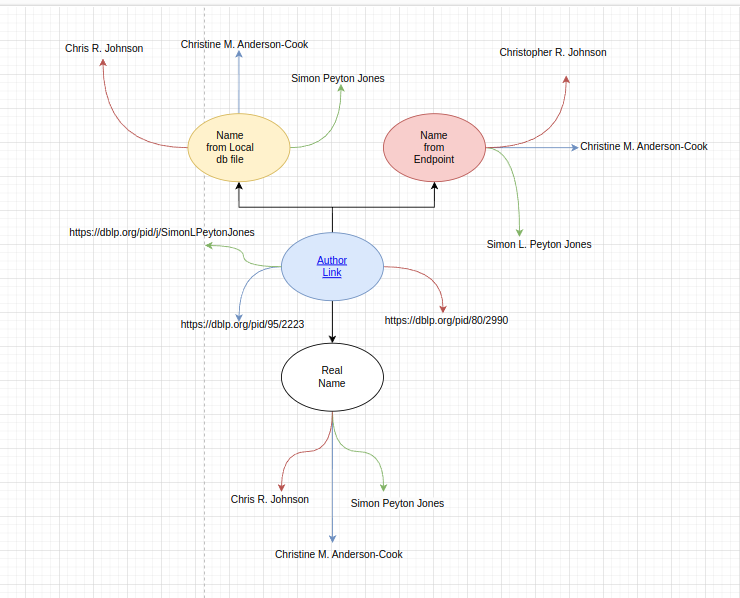}
    \caption{Data processing and cleaning.}
    \label{fig:pipeline}
\end{figure}

Fig. \ref{fig:pipeline} provides an overview of the data processing outputs from the various knowledge graphs (KGs). The data processing involved cleaning the dataset to remove noise, which included unadded parts (e.g., incomplete data fields), misspelled names, and irrelevant information such as broken or non-useful links. The knowledge graph responses were cleaned by cross-referencing returned data with authoritative sources like DBLP and SemOpenAlex to correct inconsistencies and remove redundancies. The breakdown of the cleaned data is illustrated in Equation~\ref{eq:broke-set}, which categorizes the information into distinct sets (e.g., \texttt{Authors}, \texttt{Institution}). 

An example of how the "CitedBy count" set was created from this cleaning process is shown in Fig. \ref{fig:citedby-set}, similar to how the other broken sets (e.g., \texttt{hIndex}, \texttt{i10index}) were generated. Below are the key steps in data processing:
\begin{enumerate}
    \item The datasets were transformed alphabetically according to the questions. This allowed us to identify similar structures or patterns in the questions in terms of the responses that would be returned.
    \item We compared the names returned by the \textit{author\_dblp\_uri} with those found in both the SPARQL endpoint results and the \texttt{semoa\_authors.trig} knowledge graph file.
\end{enumerate}

\begin{equation}
M = \\
\text{Breakdown sets}(D\&C) = \left\{ \begin{array}{l}
\text{List\_author\_dblp\_uri}_{\text{set}} \\
\text{Authors}_{\text{set}} \\
\text{institution}_{\text{set}} \\
\text{hIndex}_{\text{set}} \\
\text{i10index}_{\text{set}} \\
\text{acronym}_{\text{set}} \\
\text{etc}
\end{array} \right.
    \label{eq:broke-set}
\end{equation}

\subsection{Divide and Conquer Approach}
\label{divideConquer}
\begin{figure}[h!]
    \centering
    \includegraphics[width=0.6\linewidth]{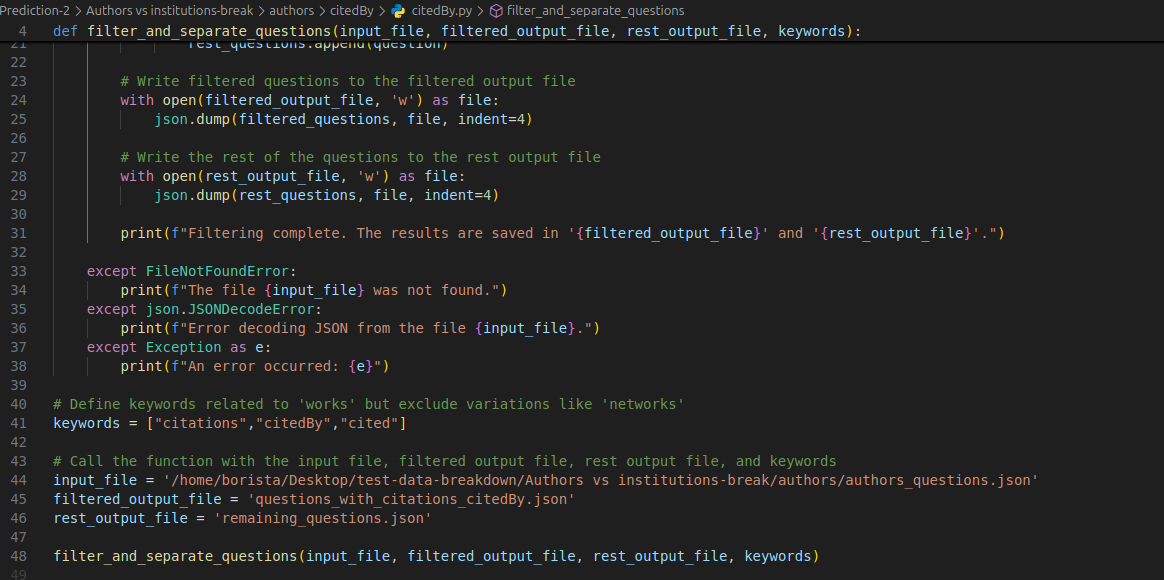}
    \caption{Example of the "CitedBy count" set creation.}
    \label{fig:citedby-set}
\end{figure}

To manage the diverse nature of the questions and data, we implemented a divide and conquer strategy:

\begin{enumerate}
    \item \textbf{Initial Data Breakdown:} The test data were first segmented based on whether the \texttt{author\_dblp\_uri} contained multiple links or a single link. This allowed us to address questions with multiple author identifiers separately from those with single identifiers.
    
    \item \textbf{Further Segmentation:} Questions were further classified into those concerning individual authors and those related to institutions. Keywords like "Organizations," "Affiliations," "Institution," etc. were used to automate this classification based on the content of the questions.

    \item \textbf{Detailed Sub-Classification:} Within the author-related questions, further subdivisions were made depending on the specific information sought. For example, we classified questions according to whether they requested publication details, citation counts, or institutional affiliations. This sub-classification relied heavily on extracting and analyzing keywords within the questions to tailor the SPARQL queries appropriately.

    As illustrated in Equation~\ref{eq:broke-set}, the dataset was broken into different sets such as \texttt{List\_author\_dblp\_uri}, \texttt{Authors}, \texttt{Institution}, \texttt{hIndex}, and more. Each set represented a component that corresponded to the key information extracted from the questions. For instance, Fig. \ref{fig:citedby-set} provides a visual example of how the "CitedBy count" set was created. This method was applied similarly to other components, ensuring a structured approach for processing and querying different knowledge graphs.
\end{enumerate}

\subsection{Data Retrieval and Aggregation}
\label{dataRetrieval}
We employed a script to generate a CSV file containing all potential responses for each question by querying the endpoints provided on the challenge's website. This CSV file included detailed author information such as names, publication counts, and institutional affiliations.

\begin{enumerate}
    \item \textbf{CSV to JSON Conversion:} The CSV file was converted to JSON format. Duplicate entries, resulting from multiple author names, were removed to ensure a clean dataset.
    \item \textbf{Merging Results:} The JSON file was then used to cross-reference and extract answers for each specific question. The answers were aggregated and merged to create a comprehensive set of responses.
    \item \textbf{Final Refinement:} The merged results were refined by integrating them with the initial general predictions and LLM-generated responses to ensure accuracy and completeness. This final step resulted in the creation of the \texttt{answers2.txt} file submitted for evaluation.
\end{enumerate}

\subsection{Large Language Model-Based Predictions}
\label{llmPredictions}
The LLM used in this challenge was BERT-base-cased-SQuAD2, a pretrained model fine-tuned on the Stanford Question Answering Dataset, downloaded from Hugging Face\footnote{\url{https://huggingface.co/deepset/bert-base-cased-squad2}}. This model was chosen without additional fine-tuning due to its capability to answer scholarly-related questions from a phrase, which aligns with our methodology for this year's Question Answering tasks. Future work will focus on fine-tuning the LLM with this year’s dataset to evaluate its performance after fine-tuning. It should be noted that, although BERT does not have an exceptionally large number of parameters compared to more recent models, it is still considered an LLM, as described in section II of the article \textit{Large Language Models: A Survey}\footnote{\url{https://arxiv.org/html/2402.06196v2}}.

After executing SPARQL queries, we used the BERT-based model \texttt{bert-base-cased-squad2} to predict responses to personal questions about authors. The context for these predictions was generated from the results of the SPARQL queries. This step was crucial for answering questions that required detailed and context-specific information. The overall LLM prediction steps are:
\begin{enumerate}
    \item \textbf{Context Generation:} The context for each question was constructed from the data retrieved through SPARQL queries.
    \item \textbf{LLM Inference:} Using the \texttt{bert-base-cased-squad2} model, we generated predictions based on the context. This model was trained on the SQuAD2 dataset to handle the intricacies of question answering with contextual information. This model was from Hugging Face\footnote{\url{https://huggingface.co/deepset/bert-base-cased-squad2}}.
    \item \textbf{Integration:} The LLM-generated responses were integrated with the initial query-based results before the final refinement stage to enhance the accuracy and completeness of the answers. The final refinement involved combining questions resolved by the Local Predictions and those resolved using LLM-based predictions, as well as the questions found in the combined output from the Divide and Conquer(D\&C) algorithms that were not present in the \textit{answers2.txt}.
    \end{enumerate}

\subsection{How questions were approached}

\begin{figure}[h!]
    \centering
    \includegraphics[width=1.0\linewidth]{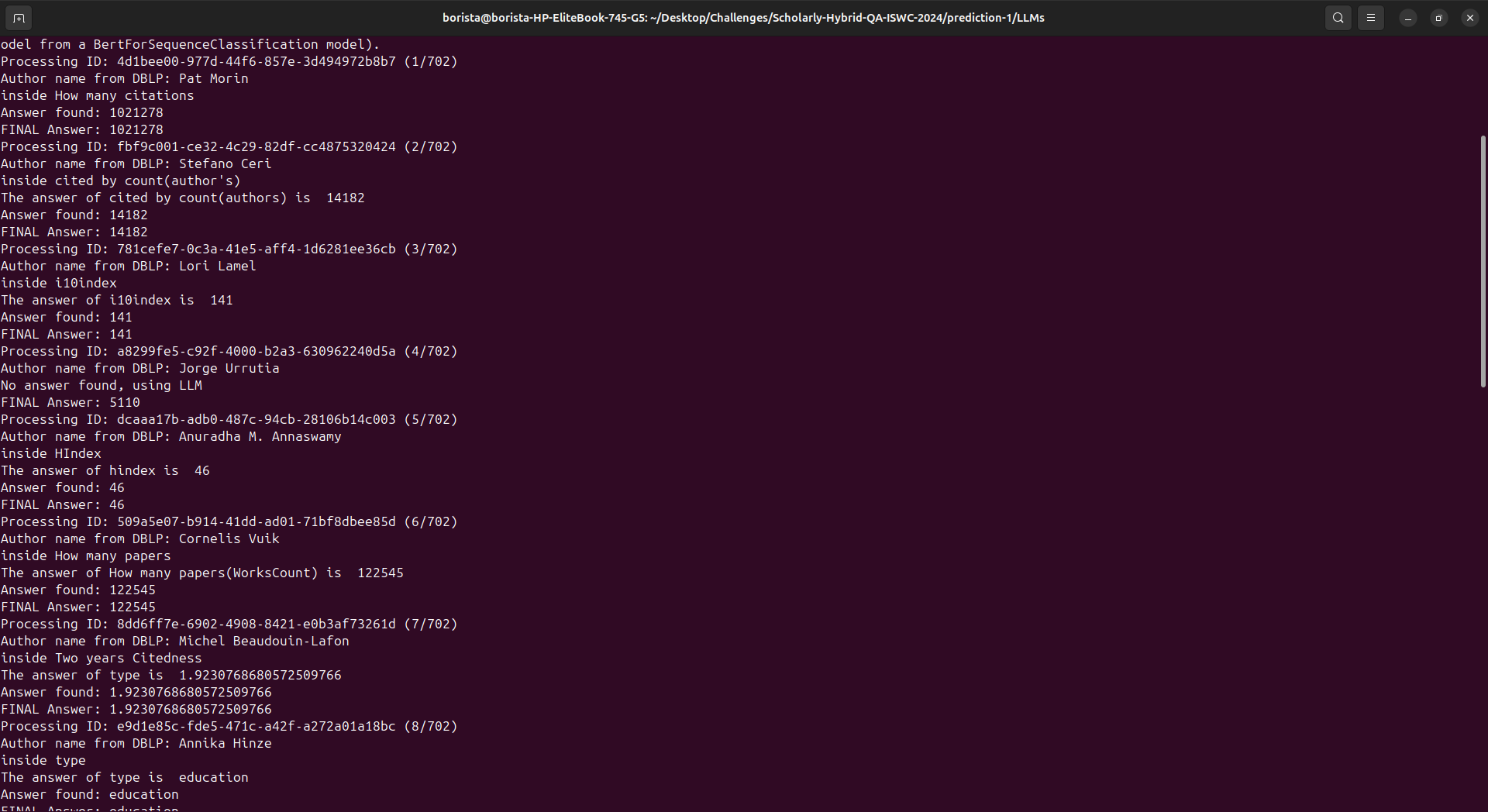}
    \caption{Sample of how questions were approached}
    \label{fig:sample-of-questions}
\end{figure}

Fig. \ref{fig:sample-of-questions} provides an overview of the data processing outputs from the various knowledge graphs (KGs). The process of answering a question typically begins by using the \texttt{author\_dblp\_link} to retrieve the author's name from the DBLP knowledge graph. Once the author names are identified, they undergo a series of SPARQL queries to extract relevant information about the author from both the DBLP and SemOpenAlex KGs.

\begin{figure}[h!]
    \centering
    \begin{minipage}{.3\textwidth}
        \centering
        \includegraphics[width=1.0\linewidth]{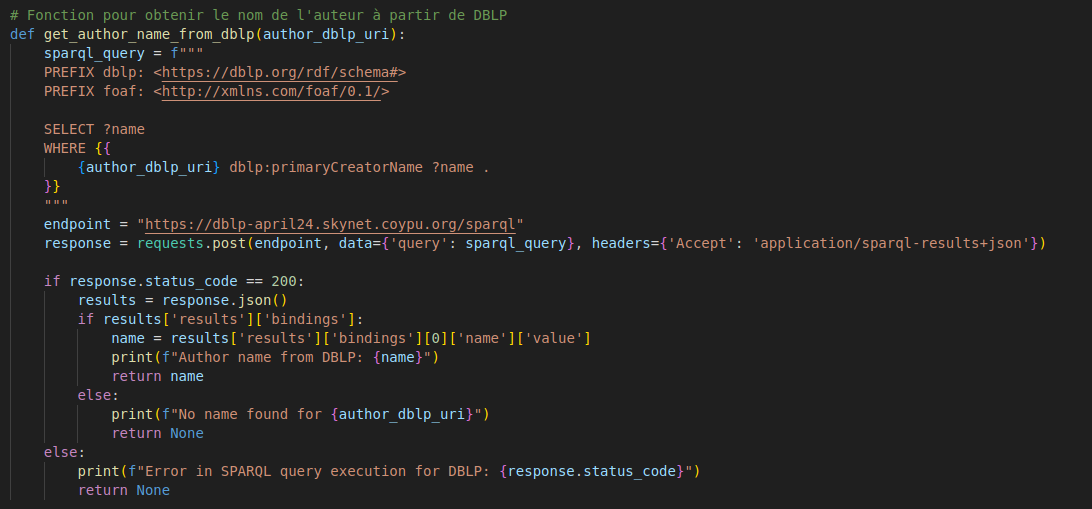}
        \caption{SPARQL query to retrieve author name}
        \label{fig:sparql-author-name-query}
    \end{minipage}\hfill
    \begin{minipage}{.3\textwidth}
        \centering
        \includegraphics[width=1.0\linewidth]{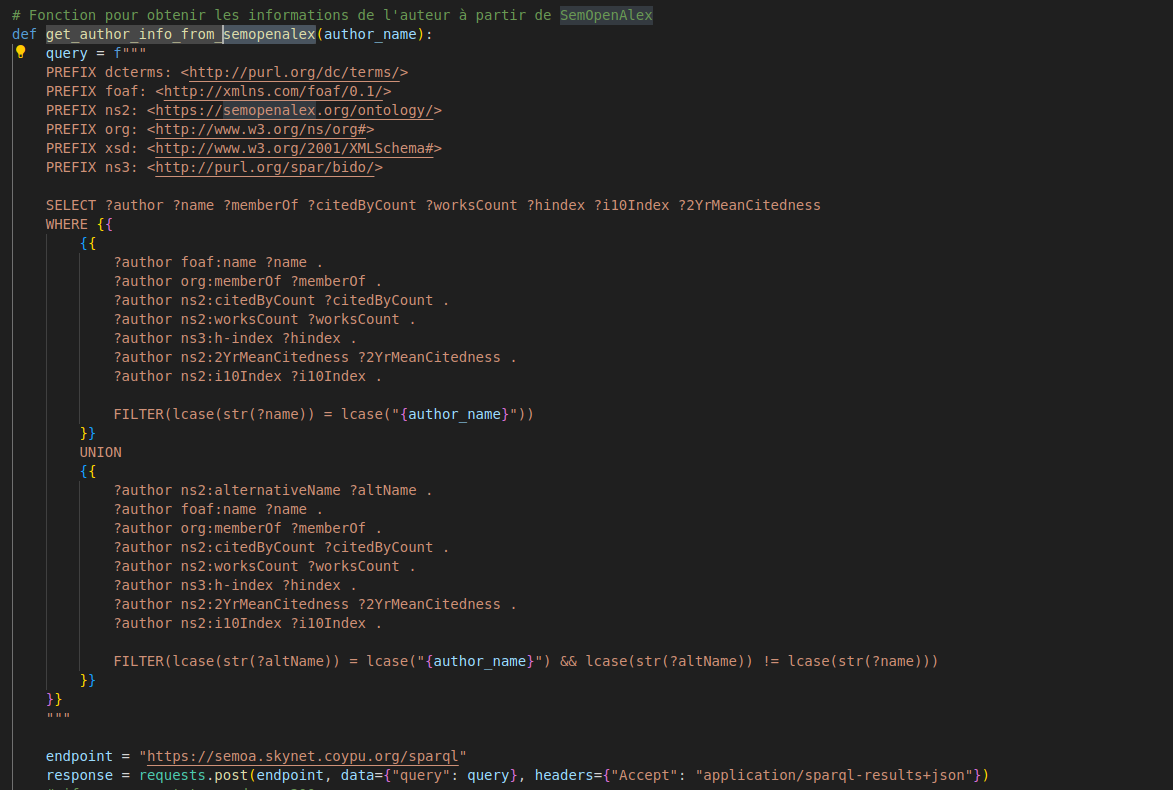}
        \caption{SPARQL query to retrieve author information}
        \label{fig:sparql-author-information-query}
    \end{minipage}\hfill
    \begin{minipage}{.3\textwidth}
        \centering
        \includegraphics[width=1.0\linewidth]{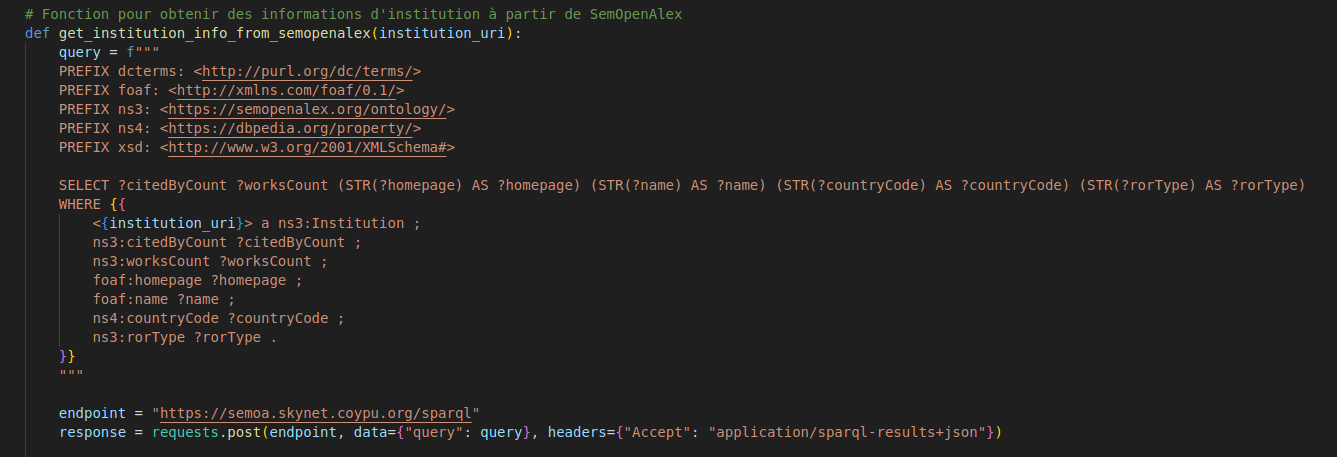}
        \caption{SPARQL query to retrieve author's institution information}
        \label{fig:sparql-author-institution-information-query}
    \end{minipage}
\end{figure}

For instance, the first query fetches the author's name using their unique DBLP link, as depicted in Fig. \ref{fig:sparql-author-name-query}. This is followed by additional queries that gather further author-specific details, such as the number of publications and citation counts, shown in Fig. \ref{fig:sparql-author-information-query}. Finally, institution-related data about the author is retrieved, as shown in Fig. \ref{fig:sparql-author-institution-information-query}.

\subsection{Evaluation and Finalization}
The evaluation of our approach was carried out by submitting the results obtained after the application on the test set to the codalab \footnote{\url{https://codalab.lisn.upsaclay.fr/competitions/19747}} provided by the organisers. The results was assessed based on Exact Match and F-score metrics.

\subsection{Experimentation Environment}
The experimentation was conducted using an HP EliteBook 745 G5 laptop equipped with an AMD Ryzen™ 5 PRO 2500U w/ Radeon™ Vega Mobile Gfx × 8 CPU, 24 GB of RAM, and a 512.0 GB SSD disk. The operating system used was Ubuntu 24.04.4 LTS.

\section{Results and Discussion}
\label{results}
\begin{figure}[h]
    \centering
    \includegraphics[width=0.6\linewidth]{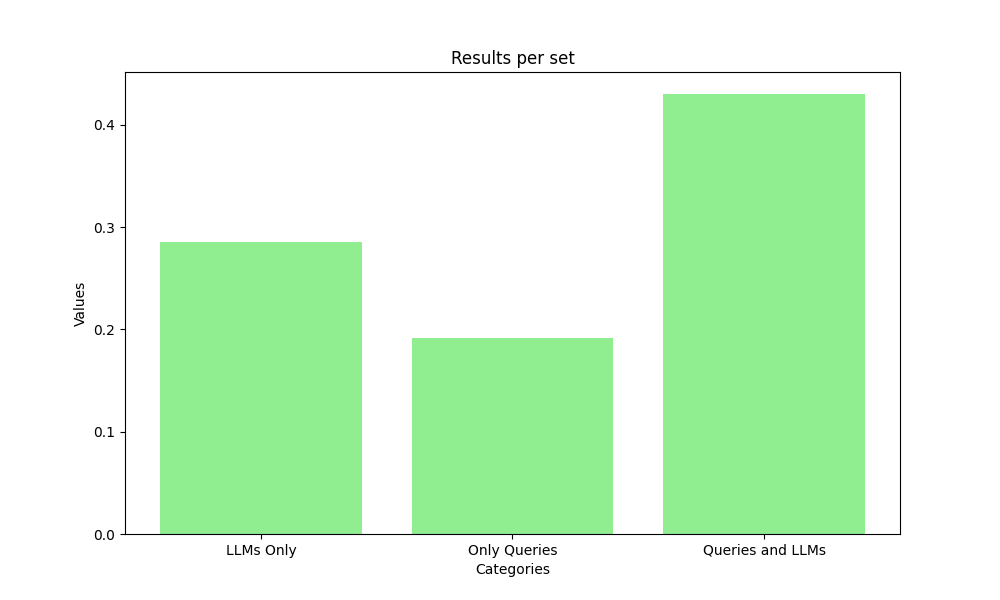}
    \caption{Performance comparism of the different(All) results of our model.}
    \label{fig:semopenalex_results}
\end{figure}

Fig. \ref{fig:semopenalex_results} presents the results obtained after applying the methodology presented in Section \ref{methodology}. It shows that the best results is obtained when the SPARQL queries are combined with the LLM for predicting responses.

During this work, we found that:
\begin{itemize}
    \item To manage complex queries on authors, institutions, affiliations and publications on the Semopenalex, we integrated SPARQL queries and LLMs prediction.
    \item the BERT-base-cased-squad2 model combined with Divide and conquer(D\&C) algorithm significantly improved the accuracy of entity and relation extraction on the DBLP KG. It should be noted that these information are needed to provide the context for the prediction by the LLMs.
    \item To handle the complete dataset, the Divide and conquer(D\&C) algorithm was employed, so as to be able to get through all the broken sets of the dataset. 
\end{itemize}

\section{Conclusion}
\label{conclusion}
In this paper, we presented a novel approach for Hybrid Question answering over Linked Data. This approach was assessed on the training and test datasets of the Scholarly Hybrid Question Answering over Linked Data (QALD) Challenge 2024. We found that the integration of SPARQL queries with LLM-based predictions offers a robust solution for Question Answering over diverse scholarly data sources. Our approach demonstrated significant improvements in handling complex queries and providing accurate responses. Despite the results obtained, there were several challenges, particularly in handling the large and complex nature of the SemOpenAlex and DBLP datasets. Future work will focus on improving the model's ability to generalize across different types of scholarly data and incorporating more sophisticated rule-based systems on the one hand. On the other hand, we will focus on refining the methodology and exploring additional enhancements to further improve the system's performance.

\section{Online Resources}

The source code for this project is available via
\begin{itemize}
\item \href{https://github.com/FOMUBAD-BORISTA-FONDI/Scholarly-Hybrid-QA-ISWC-2024}{GitHub}
\end{itemize}


\begin{thebibliography}{99}

\bibitem{AjVtBf2020AtP}
A. Jiomekong, V. Tsague, B. Foko, U. Melie and G. Camara, "Towards an approach based on knowledge graph refinement for answer type prediction," 2020.
\bibitem{AjVtBf2019AtP}
A. Jiomekong, G Camara and M. Tchuente "Extracting ontological knowledge from java source code using hidden markov models," \textit{Open Computer Science,} Vol. 9, no. 1, pp. 181-199, 2019.
\bibitem{mao2022hybridqa}
Y. Mao et al., "HybridQA: A Dataset of Multi-Hop Question Answering over Tabular and Textual Data," in \textit{Proceedings of the 2022 Conference on Empirical Methods in Natural Language Processing}, 2022, pp. 5130-5141.
\bibitem{qald2023}
J. Smith and A. Doe, \textit{Scholarly QALD 2023: Enhancing Question Answering over Knowledge Graphs}, in \textit{Proceedings of the Scholarly QALD 2023 Workshop}, 2023.
\bibitem{iswc2024}
International Semantic Web Conference (ISWC) 2024. Available at: \url{https://iswc2024.semanticweb.org}
\bibitem{qald2023}
J. Smith and A. Doe, \textit{Scholarly QALD 2023: Enhancing Question Answering over Knowledge Graphs}, in \textit{Proceedings of the Scholarly QALD 2023 Workshop}, 2023.
\bibitem{Wikipedia}
Wikipedia contributors, \textit{Wikipedia-Based Scholarly Text}, retrieved from \url{https://drive.google.com/file/d/1ISxvb4q1TxcYRDWlyG-KalInSOeZqpyI/view?usp=drive_link}, 2024.
\bibitem{Russell1999EM}
S. Russell and P. Norvig, \textit{Artificial Intelligence: A Modern Approach}, 2nd ed., Prentice Hall, 1999.
\bibitem{divide_conquer_ref}
D. S. Johnson and C. H. Papadimitriou, "On the Performance of Divide and Conquer Algorithms," in \textit{Journal of Computer and System Sciences}, vol. 19, no. 2, pp. 127-149, 1981.
\bibitem{devlin2018bert}
J. Devlin, M. Chang, K. Lee, and K. Toutanova, "BERT: Pre-training of Deep Bidirectional Transformers for Language Understanding," in \textit{Proceedings of the 2019 Conference of the North American Chapter of the Association for Computational Linguistics: Human Language Technologies}, vol. 1, pp. 4171-4186, 2019.
\end{thebibliography}
\end{document}